\documentstyle[aps,preprint]{revtex}
\begin{document}
\title{Randomly Charged Polymers: An Exact Enumeration Study}
\author{Yacov Kantor}
\address{School of Physics and Astronomy, Tel Aviv University,
Tel Aviv 69 978, Israel}
\author{Mehran Kardar}
\address{Department of Physics, Massachusetts Institute of
Technology, Cambridge, MA 02139, U.S.A.}
\date{\today}
\maketitle
\tightenlines
\begin{abstract}
We perform an exact enumeration study of
polymers formed from a (quenched) random sequence
of charged monomers $\pm q_0$. Such polymers, known
as polyampholytes, are compact when completely neutral
and expanded when highly charged. Our exhaustive search
included all spatial conformations and quenched sequences for
up to 12--step (13--site) walks. We investigate the behavior
of the polymer as a function of its overall excess
charge $Q$, and temperature $T$.
At low temperatures there is a phase transition from compact
to extended configurations when the charge exceeds
$Q_c\approx q_0 \sqrt{N}$.
There are also indications of a transition for small $Q$
between two compact states on varying temperature.
Numerical estimates are provided for the condensation
energy, surface tension, and the critical exponent $\nu$.
\end{abstract}
\section{Introduction}  \label{secintro}

Given their ubiquity in nature, long chain macromolecules
have been the subject of considerable study.
Whereas there is now a reasonably
firm basis for understanding the physical properties of
homopolymers\cite{rPolGen}, considerably less is known about
the heteropolymers of biological significance. From a biologist's
perspective, it is the specific properties of a particular
molecule that are of interest. After all the genetic information
is coded by very specific sequences of nucleic acids, which are
in turn translated to the chain of amino acids forming a
protein\cite{rgen}. The energy of the polymer is determined by
the van der Waals, hydrogen bonding, hydrophobic/hydrophilic,
and Coulomb interactions between its constituent amino acids.
In accord to these interactions, the protein folds into a
specific shape that is responsible for its activity.
Given the large number of monomers making up such chains,
and the complexity of their interactions, finding the
configuration of a particular molecule is a formidable task.
By contrast, a physicist's approach is to sacrifice the
specificity, in the hope of gleaning some more general information
from simplified models\cite{rstein}. There are in fact a number
of statistical descriptions of {\it ensembles} of molecules composed of a
random linear sequence of elements with a variety of interactions
that determine their final shapes\cite{rGOa}.
These simple models of heteropolymers are of additional interest as
examples of disordered systems with connections to spin--glasses
\cite{spinglass}, with the advantage of faster
relaxation \cite{rExT,anaka}.

There are a number of recent experimental studies of
solutions\cite{rcand} and gels\cite{rExT,anaka} of
polymers that incorporate randomly charged groups.
As statistical approaches only provide general
descriptions of such heteropolymers, we focus on
simple models which include the essential ingredients.
The overall size and shape of a polymer with charged groups
is most likely controlled by the Coulomb interactions
that are the strongest and with the longest range. We shall
consider the typical properties of a model
{\it polyampholyte} (PA)\cite{rPA}:
a flexible chain in which each of the $N$ monomers
has a fixed charge $\pm q_0$ selected from a well defined
ensemble of quenches. The polymer has a characteristic microscopic
length $a$ (such as range of the excluded--volume
interaction, or nearest neighbor distance along the chain).
In the numerical studies we further simplify the
model by considering only self--avoiding walk (SAW)
configurations on a cubic lattice with lattice constant $a$.

The long range nature of the Coulomb interactions, combined
with the randomness of the charge sequence, produces
effects quite distinct from systems with short range
interactions. In Section \ref{secgend} we use the knowledge
accumulated in
previous studies\cite{rKK,rHJ,rKLKprl,rKKLpre,KKshort,KKlong}
to explore the phase diagrams of quenched PAs
in $d$ dimensions. In particular, we show that for $d>4$,
the behavior of PAs is similar to that of random chains
with short--range interactions, while for $d<4$ the spatial
conformations of a PA strongly depend on its
excess charge $Q\equiv\sum_{i=1}^Nq_i$. In every space
dimension $d<4$, there is a critical charge $Q_R$
such that PAs with $Q>Q_R$ cannot form a compact state.
The probability of a randomly charged PA to have such an
excess charge depends on both $d$ and its length.
In the $N\rightarrow\infty$
limit the excess charge will always (i.e. with probability
1) be ``small'' for $d>3$ and ``big'' for $d<3$. Thus
investigation of the ``borderline'' three--dimensional case
provides valuable insight into the behavior of the system
in general space dimensions.

In Section \ref{secgen} we summarize previous results
for PAs in $d=3$: Analytical arguments and Monte
Carlo (MC) studies indicate that the PA undergoes a transition
from a dense (``globular'') to a strongly
stretched configuration as $Q$ exceeds $Q_c\approx q_0N^{1/2}$.
The MC simulations\cite{rKLKprl,rKKLpre,KKshort,KKlong}
were performed for polymer sizes up to $N=128$ and in a wide
range of temperatures. They, however, could not provide
information on the energy spectrum of PAs, and on very low
temperature properties. In this work we undertake a complete
enumeration study of PAs for all possible quenches up to
$N=13$, and are thus able to present very detailed results
regarding energetics and spatial conformations of short PAs.
The details of the enumeration procedure are explained
in Section \ref{secenum}, while the results are described
in Sections \ref{secenspec} and \ref{secshape}. The majority
of these results add further support to the predictions
of MC studies, and provide some details which could not be
measured by MC (e.g., density of states, condensation
energy, and surface tension in the globular phase). We also
find some indication that PAs with small $Q$ may undergo a
phase transition between two dense states.
No signs of this transition could be detected in the MC studies,
because it occurs at temperatures too low
for that procedure to equilibrate.

\section{Polyampholyte Phenomenology}\label{secgend}

It is helpful to view the problem in the more general context
of a variable space dimension $d$. Let us consider a continuum
limit in which configurations of the PA are described by a
function $\vec{r}(x)$. The continuous index $x$ is used to
label the monomers along the chain, while $\vec{r}$ is the
position of the monomer in $d$--dimensional embedding space.
The corresponding probabilities of these configurations are
governed by the Boltzmann weights of an effective Hamiltonian,
\begin{eqnarray}\label{continuumH}
{{\cal H}[q(x)]\over T }& = &
{K\over2}\int dx\left({d\vec{r}\over dx}\right)^2
+{v\over2}\int dxdx'\delta^d(\vec{r}(x)-\vec{r}(x'))
\nonumber\\
        & & +{1\over 2T}\int dxdx'{q(x)q(x')\over
|\vec{r}(x)-\vec{r}(x')|^{d-2}} \nonumber \\
& \equiv &  H_0+H_v+H_q\ .
\end{eqnarray}
In this equation $H_0$ represents the entropic properties
of the connected chain (ideal polymer), $H_v$ is the
continuum description of the excluded volume interactions,
while $H_q$ represents the $d$--dimensional electrostatic
energy. For each PA, there is a specific (quenched) function
$q(x)$ representing the charges along the chain.
(In this work we set $k_B=1$ and measure $T$ in energy units.)

In the simplest ensemble of quenches, each monomer takes a
charge $\pm q_0$ independent of all the others; i.e.
$\overline{q_iq_j}=\delta_{ij}q_0^2$, where the overline
indicates averaging over quenches.
While the average charge of such PAs is zero, a ``typical''
sequence has an excess charge of about $\pm Q_c$,
with $Q_c\equiv q_0N^{1/2}$. This statement, as well
as the definition of $Q_c$, are unrelated to the embedding
dimension $d$. However, the importance of charge fluctuations
(both for the overall polymer, or large segments of it)
does depend on the space dimension.
The electrostatic energy of the excess charge,
spread over the characteristic size of an ideal polymer
($R\propto N^{1/2}$), grows as $Q^2/R^{d-2}\sim N^{(4-d)/2}$.
This simple dimensional argument shows that for $d>4$ weak
electrostatic interactions are irrelevant. (The excluded volume
effects are also irrelevant in $d>4$.)
Thus, at high temperatures the PA behaves as an ideal polymer
with an entropy--dominated free energy of the order of $-NT$.
However, on lowering temperature it collapses
into a dense state, taking advantage
of a condensation energy of the order of $-Nq_0^2/a^{d-2}$.
This collapse is similar to the well known $\theta$--transition
of polymers with short range interactions and will be
discussed later in this Section.

For $d<4$, electrostatic interactions are relevant and the
high temperature phase is no longer a regular self--avoiding
walk. At high temperatures the behavior of the polymer can be
studied perturbatively. For the above ensemble of uncorrelated
charges, the lowest order ($1/T$) correction to the
quench--averaged $R_g^2$ vanishes\cite{rKLKprl,rKKLpre}.
However, if we restrict the ensemble of quenches to sequences
with a fixed overall excess charge of $Q$, there is a lowest
order correction term proportional to $Q-Q_c$. Thus PAs with $Q$
less than $Q_c$ contract while those with larger charges
expand. This trend appears in any space dimension $d$, and is
indicated by the vertical line at the top of Fig.~\ref{FigA}.
It should be noted that restricting the ensemble to
yield fixed $Q$, slightly modifies the quench--averaged
charge--charge correlations. In
particular, the two--point correlation function becomes
$\overline{q_iq_j}=(Q^2-Q_c^2)/N^2$ for $i\ne j$.
This small (order of $1/N$) correction to the correlation
function may cause a significant change in $R_g^2$ due to the
long range nature of the Coulomb interaction.

The above discussion can be extended to PAs with short--range
correlations along the sequence: If neighboring charges satisfy
$\overline{q_iq_{i+1}}=q_0^2\lambda$\cite{rwittmer}
where $-1<\lambda<1$, with no further restrictions,
then $\overline{q_iq_j}=q_0^2\lambda^{|j-i|}$. The resulting
ensembles continuously interpolate between the deterministic
extremes of an alternating sequence ($\lambda=-1$) and a
uniformly charged polyelectrolyte ($\lambda=1$). As in the case
of uncorrelated charges, we can impose an additional
constraint on the overall charge, resulting in correlations
\begin{equation}\label{qqcorr}
\overline{q_iq_j}=q_0^2\lambda^{|j-i|}+
(Q^2-Q_c^2(\lambda))/N^2\ ,
\end{equation}
where $Q_c^2(\lambda)=q_0^2N(1+\lambda)/(1-\lambda)$.
We note that the variance of $Q$ in such a correlated
sequence also becomes $q_0^2N(1+\lambda)/(1-\lambda)$.
Thus the proportion of quenches with $Q$
above or below $Q_c(\lambda)$ is independent of
$\lambda$. All the results for uncorrelated sequences
remain valid if we substitute $Q_c(\lambda)$ for $Q_c$.
As $|\lambda|\rightarrow 1$, the behavior of the PA crosses over
from that of a random sequence to the deterministic
(alternating or homogeneous) one. However, the crossover occurs
only for $|\lambda|-1\approx {\cal O}(1/N)$. As typical of the qualitative
behavior outside this narrow interval,
we concentrate on the uncorrelated case of $\lambda=0$.

A short distance cutoff $a$, such as the range of
the excluded volume interaction, introduces
a temperature scale $q_0^2/a^{d-2}$. For $d>4$
the electrostatic interactions in random PAs are
effectively short--ranged. Previous results on a random short
range interaction model\cite{rsrim} (RSRIM) in $d=3$
indicate that, as long as the positive and negative
charges are approximately balanced, the polymer
assumes spatial conformations where the interactions
are predominantly attractive. To maximize this attraction,
the chain undergoes a transition from an expanded to a
collapsed (dense) state at a $\theta$--transition. For truly
short range interactions, the $\theta$--transition disappears
only for a rather strong charge imbalance of $Q\sim N$.
Even if as a result of the relevant Coulomb interactions in
$d<4$, the high temperature phase of uncorrelated PAs turns
out to be compact, we cannot exclude the possibility of
a transition into another dense (possibly glassy) state when $T$
decreases below a critical $\theta_E$. Such a potential
``$\theta_E$--transition'', indicated by the horizontal dashed
line in Fig.~\ref{FigA}, must be different from a regular
collapse since the lower density phase is not a
self--avoiding walk. The compact phase can also be destroyed
by increasing the net charge as described
in the following paragraph.

A dense globular PA droplet of radius $R\sim aN^{1/d}$
has a surface energy of $\gamma R^{d-1}$, where
$\gamma\approx q_0^2/a^{2d-3}$ is the surface tension.
For small $Q$, the
surface tension keeps the PA in an approximately spherical
shape. However, as shown in Appendix~\ref{secrayleigh},
at sufficiently large $Q$ electrostatic forces
destabilize the droplet. Comparing the electrostatic ($\propto Q^2/R^{d-2}$)
and surface energies indicates that the droplet shape
is controlled by the parameter $\alpha=Q^2/Q_R^2$,
where $Q_R\approx q_0N^{1-3/2d}$ is the {\it Rayleigh
charge}. For a large enough $\alpha$ a spherical
shape is unstable (a charged liquid droplet disintegrates).
The Rayleigh charge in $d=3$ is proportional to
$Q_c=q_0\sqrt{N}$, while for $d>3$ ($d<3$) it increases faster
(slower) than $Q_c$. The solid vertical line at the
bottom of Fig.~\ref{FigA} shows the position of this instability
in $d=3$. Clearly, any $\theta_E$--transition (if at all
present) must also terminate at $Q_R$. Only a negligible
fraction of random quenches in $d>3$ have $Q$ exceeding $Q_R$,
and thus a typical PA is a spherical droplet at low $T$.
Conversely, in $d<3$ almost all PAs have charges larger
than $Q_R$ and the dense phase does not exist.
The borderline case of $d=3$, where a finite fraction of
PAs have $Q$ exceeding $Q_R$, is the most controversial:
An analogy with uniformly charged polyelectrolytes\cite{rPVdG}
suggests\cite{rKK} that the
PA is fully stretched ($\nu=1$) in this case\cite{rKK}.
By contrast, a Debye--H\"uckel inspired theory\cite{rHJ}
predicts that low--$T$ configurations are compact.
Partial resolution of this contradiction comes from the
observation\cite{rKLKprl,rKKLpre} that PAs in $d=3$ are
extremely sensitive to the excess charge $Q$. In the
following Section we shall briefly review the main features
of three--dimensional PAs obtained by MC
simulations\cite{KKshort,KKlong}.

\section{Monte Carlo Results in three dimensions}\label{secgen}

Numerical simulations are performed on a discretized version of
Eq.~(\ref{continuumH}). Configurations of a polymer are
specified by listing the position vectors $\{{\bf r}_i\}$
($i=1,\dots,N$) of its monomers. The shape and spatial
extent of the polymer are then characterized by the tensor,
\begin{equation}\label{shapetensor}
{\cal S}_{\mu\nu}={1\over N}\sum_{i=1}^Nr_{i\mu}r_{i\nu}
-{1\over N^2}\sum_{i=1}^Nr_{i\mu}\sum_{j=1}^Nr_{j\nu}\ ,
\end{equation}
with the greek indices labeling the various components.
Thermal averages of the eigenvalues
$\lambda_1>\lambda_2>\lambda_3$ of this tensor
(sometimes referred to as moments of inertia) are used to
describe the mean size and shape; their sum
is the squared radius of gyration, $R_g^2={\rm tr}{\cal S}$. Since we are
dealing
with sequences of quenched disorder, these quantities must
also be averaged over different realizations of $\{q_i\}$.
In three dimensions, uniform uncharged polymers in good
solvents are swollen; their  $R_g$ scaling as  $N^\nu$ with
$\nu=0.588$ as in self--avoiding walks. Polymers in poor
solvents are ``compact'', i.e. described by $\nu={1/3}$.

In previous work\cite{KKshort,KKlong} we used Monte Carlo (MC)
simulations (along with analytical arguments) to establish the
following properties for PAs immersed in a good solvent:

(a) The radius of gyration strongly depends on the total
excess charge $Q$, and is weakly influenced by other details
of the random sequence.

(b) A $1/T$--expansion indicates that the size of a PA tends to
decrease upon lowering temperature if $Q$ is less than a
critical charge $Q_c\equiv q_0N^{1/2}$, and increases otherwise.
This behavior is confirmed by MC simulations.

(c) At low temperatures, neutral polymers ($Q=0$) are compact
in the sense that their spatial extent in any direction grows
as $aN^{1/3}$, where $a$ is a microscopic length scale.

(d) The low--$T$ size of the PA exhibits a sharp dependence on
its charge:  $R_g$ is almost independent of $Q$ for
$Q<q_0\sqrt{N}$, and grows rapidly beyond this point. This increase becomes
sharper as the temperature is lowered,
or as the length of the chain is made longer.

We interpret the low temperature results by an analogy to the
behavior of a charged liquid drop. The energy (or rather the
quench averaged free energy) of the PA is phenomenologically
related to its shape by
\begin{equation}\label{edefenerg}
E=-\epsilon_cN+\gamma S + {b Q^2\over R}.
\end{equation}
The first term is a condensation energy proportional to the
volume (assumed compact), the second term is proportional to the
surface area $S$ (with a surface tension $\gamma$), while the
third term represents the long range part of the electrostatic
energy due to an excess charge $Q$ ($b$ is a dimensionless
constant of order unity). The optimal shape is obtained by
minimizing the overall energy. The first term is the same for
all compact shapes, while the competition between the surface
and electrostatic energies is controlled by the dimensionless
parameter
\begin{equation}
\alpha\equiv {Q^2\over 16\pi R^3\gamma}=
{Q^2\over 12V\gamma}\equiv {Q^2\over Q_R^2}\ .
\end{equation}
Here $R$ and $V$ are the radius and volume of a spherical drop
of $N$ particles, and we have defined the {\it Rayleigh charge}
$Q_R$. (See Eq.~(\ref{eQR}) of the Appendix for the definition of
$Q_R$ in a general dimension $d$.) The dimensionless parameter
$\alpha$ controls
the shape of a charged drop: A spherical drop becomes unstable
and splits into two equal droplets for $\alpha>0.3$. We argued
in Ref.\cite{KKshort,KKlong} that the quenched PA has a similar
instability at the vertical line in the bottom of
Fig.~\ref{FigA}: Charged beyond a critical $\alpha$, the PA
splits to form a {\it necklace} of blobs connected by strands.
{}From the definition of $Q_R$ it is clear that it is proportional
to $Q_c$; the dimensionless prefactor relating the two depends
on $\gamma$ and is estimated in this work.

(e) While confirming the general features of Fig.~\ref{FigA},
the MC simulations provide no indication of the suggested
$\theta_E$ transition. However, these simulations are not
reliable at very low temperatures\cite{tcompare} due to
the slowing--down of the equilibration process.

All the above results were obtained by MC simulations for PAs of between 16 and
128 monomers. Since the MC procedure does not
provide good equilibration at low $T$, we could not determine
the properties of the ground states (although some conclusions
were drawn from the low--$T$ data).  Systems with Coulomb
interactions are particularly poorly equilibrated, even at
densities of only 10\% the maximal value. To remedy these
difficulties, in this work we resorted to complete enumeration
of all possible spatial conformations, and all possible quenched charge
sequences. While such an approach enables us to obtain
exact results, and detailed information not available in MC
studies, the immensity of configuration space restricts the
calculation to chains of at most 12 steps (13 monomers).

\section{The Enumeration Procedure}\label{secenum}

We considered self--avoiding walks (SAWs) on a simple cubic lattice
of spacing $a$. An $L$--step SAW has $N=L+1$ sites (atoms), and
the randomly charged polymer is defined by assigning a fixed sequence
of charges ($q_i=\pm q_0$) to its monomers. The charge
sequence is considered to be quenched, i.e. it remains unchanged
when the spatial conformation of the walk changes. The energy
of any particular configuration is given by
$U=\sum_{i<j}q_iq_j/|\vec{r}_i-\vec{r}_j|$, where $\vec{r}_i$ is
the position of the $i$th atom on the three--dimensional lattice.
Thermal averages of various quantities are calculated by summing over all
conformations with the Boltzmann weights of this energy. The resulting
average is quench specific. We then obtain quenched averages by summing
over all possible realizations of the sequence, possibly with certain
restrictions such as on the total excess charge $Q\equiv\sum_iq_i$.

Our calculation consists of three steps: (a) Generating lists of all
spatial conformations and quench sequences; (b) Using the lists to
calculate various thermodynamic quantities for each quenched configuration;
(c) Averaging of results over restricted
ensembles of quenches, and analyzing the data.
Since the calculational procedure is extremely time consuming, we
used precalculated lists of all SAWs, and of all possible sequences
of charges along the chains. Other programs then used
these two lists as input. Since the first list of all
$L$--step SAWs with $3\leq L\leq 12$ is extremely large, we tried
to reduce it by including
only ``truly different'' configurations and listing their degeneracies.
As the actual position of a walk in  space is not important,
we disregard it and only give the {\it directions} of the $L$ steps.
As the energy of a configuration is independent of its overall
orientation, we assume that the first step is taken in the $+x$ direction.
The above trivial symmetries are not included in our counting;
e.g. we assign a completely straight line the degeneracy of $m=1$.
All SAWs, except for the straight line, have a four fold degeneracy related
to rotation around the direction of the initial step. We shall therefore assume
that the first step which is not in $+x$ direction is taken along the $+y$
axis, and attribute a degeneracy $m=4$ to all walks which are not
straight lines. Every non--planar walk has an additional degeneracy due to
reflection in the $x-y$ plane, leading to a total degeneracy of $m=8$. Thus
the list of all 12--step SAWs consists of the directions of 11 successive
steps, along with a degeneracy factor. The total number
of 4,162,866 chains consists of
one straight line ($m=1$), 40,616 planar SAWs ($m=4$), and
4,122,249 non--planar SAWs ($m=8$). Accounting for degeneracies
reduces the length of the list and the time needed to calculate
various quantities by almost a factor of eight. Some chains possess
additional symmetries; e.g., by inverting the sequence of steps
we may get some other SAW in the list. We did not take
advantage of this symmetry because the distribution of quenched charges
on the chain is not necessarily symmetric under interchange of its two
ends. In any case, we use the end to end exchange symmetry in the
listing of all possible quenches.

A second input list contains all possible charge sequences; each ``quench"
for an $L$--step walk has $N=L+1$ charges. Since the energy is unchanged
by reversing the signs of all charges, we considered only configurations
in which the total charge is $Q\geq0$. This shortens the list by almost
a factor of 2. The majority of quenches are not symmetric under order
reversal, i.e. the sequence does not coincide with
itself when listed backwards. Since we are exploring all spatial
configurations,
without accounting for the end--reversal symmetry mentioned
in the previous paragraph, the list can be reduced by almost another
factor of 2 by considering only one of each pair of such
quenches (keeping track of the degeneracy). After accounting for both
charge and sequence reversal symmetries, the list for $L=12$ ($N=13$)
has only 2080 entries.

For any sequence, the number of computer operations required to calculate
the energy of a single configuration grows as $L^2$. The total number of
SAWs grows as $z^LL^x$, where $z\approx 4.68$ is the effective coordination
number for SAWs on a cubic lattice, and $x\approx0.2$. The number
of ``quenches" grows
as $2^L$. These factors limit the size of chains which can be investigated
to $L=12$ steps. At this $L$ we needed 4 weeks of CPU time on a Silicon
Graphics R4000 workstation. An increase of $L$ by a single unit multiplies
the calculation time by an order of magnitude. Thus it is impractical to
employ our procedure for chains that are much longer than 12 steps.
If, instead of completely enumerating all possible charge configurations
we confine ourselves to sampling a few hundred quenches, the calculations
can be extended to $L=13$, but not much further.

The order in which the calculations were performed is as follows: for
each SAW configuration we calculated the radius of gyration, squared
end--to--end distance, {\it and} the energies of all possible quenched
charge configurations along the backbone. These energies were then used
to update histogram tables (a separate histogram of possible energies
for each quench). Due to the long range nature of the Coulomb interaction,
the allowed energies form almost a continuous spectrum which we discretized
in units of $0.1q^2_0/a$. This discretization was sufficient to accurately
reproduce properties of the system on the temperature scales of interest.
(For $L<10$, we used a finer division of the histograms to verify that
the discretization process does not distort calculation of such properties
as the specific heat, except at extremely low temperatures.)
In addition to histograms, we also collected data about the energy, the
radius of gyration, and the end--to--end distance at the ground state of
each quench.

The issue of the multiplicity of the ground state is of much interest,
and hotly debated in the context of models of proteins\cite{groundstate}. In
the
presence of Coulomb interactions, due to the quasi--continuous nature
of the energy spectrum, the ground state is almost never degenerate (except
for the trivial degeneracies mentioned above). It is quite likely that, for
sufficiently long polymers and specific quenches, there may be exactly
degenerate ground states which are not related by symmetry operations.
However, for $L\le12$ such cases are extremely rare. Moreover, the distance
between the ground state and the second--lowest energy state remains
of the order of $0.1q^2_0/a$ for all $L$s in our calculation,
even though at higher energies the densities of the states increase very
rapidly with $L$.

We used this data to obtain averages over quenches
(with or without a constraint on the net charge).
It should be mentioned that creation of histograms,
as well as calculation of the thermal averages, required
the correct accounting of degeneracies of spatial
conformations, while averages over quenches
needed proper care of the sequence degeneracies.
For few selected quenches we also performed a
calculation of the density of states as a function
of two variables, the energy and the squared radius
of gyration. Due to large amount of data, we could not
do such detailed studies for all possible quenches.

Fig.~\ref{FigB} depicts ground states of
four cases of quenched charges with different excess charges $Q$.
It provides qualitative support for the conclusions previously
obtained for MC simulations: Fig.~\ref{FigB}a depicts the ground
state configuration of an almost neutral PA which is quite compact.
The PA in Fig.~\ref{FigB}b  has $Q$ slightly smaller than $Q_c$;
while the configuration is still compact we see the beginning of a
stretching. Figs.\ \ref{FigB}c and \ref{FigB}d show strongly
stretched configurations in cases where $Q$ exceeds $Q_c$.
In the following Sections we will quantify this qualitative
observations.

As a by--product of the above procedure we also obtained similar
data for the model with short--range interactions: In the random
short range interaction model (RSRIM), a quenched sequence of
dimensionless charges $q_i=\pm1$ is defined along the chain.
The interaction energy is $U=\sum_{i<j}V_{ij}$, where
$V_{ij}=v_0q_iq_j$ if $|\vec{r}_i-\vec{r}_j|=a$, and $=0$,
otherwise. While we shall compare and contrast several
properties of short and long range models in this paper, detailed results
for RSRIM can be found in Ref.\cite{rsrim}.
The additional data were gathered without a
substantial increase in the total execution time of the
programs. There are a few minor differences in the data
collection process in the two models: (a) As the energies
of the RSRIM are naturally discretized,
the resulting histograms are exact. (b) The ground state
of most quenches is highly degenerate. This
required keeping track of the degeneracy, and obtaining the
$R_g^2$ in the ground state as an average lowest energy
configurations.

\section{Energy Spectra of Quenched Polyampholytes} \label{secenspec}

We begin our analysis by testing the validity of
Eq.~(\ref{edefenerg}) for the ground states of the
polymers. Obviously, the exact value of each ground state
energy depends on the details of the charge sequence.
However, Eq.~(\ref{edefenerg})  implies that the effect of the
overall charge can be (approximately) separated;
the remaining parts of the energy depending only weakly
on the details of the sequence.
The basic energy unit of our model is $q_0^2/a$, and
a useful system for comparison is
the regular crystal formed by alternating charges
(the ``sodium chloride'' structure). The condensation energy
per atom of such a crystal ($\epsilon_{c0}=0.8738q_0^2/a$)
is much smaller than the interaction energy
per atom between the nearest neighbors ($3q_0^2/a$).
This demonstrates the importance of the long--range Coulomb interaction:
although the system is locally neutral, the ground state energy
depends on an extended neighborhood. Similarly, the surface tension
$\gamma_0=0.03q_0^2/a^3$ of the crystal is quite small.

Our first observation is that, for a fixed $Q$, the ground state energy is
quite insensitive to the details of the sequence:
Fig.~\ref{FigC} depicts the ground state energies of {\it all} 2080
possible quenches for 12--step (13--atom) chains.
(The horizontal axis represents an arbitrary numbering of the quenches.)
The energies are clearly separated into 7 bands, corresponding to
excess charges of  $Q=1$, 3, 5, $\cdots$, 13. (There is only
one quench with $Q=13$.) While each band has a finite width, we
see that the energy of a PA can be determined rather accurately
by only specifying its net charge $Q$! {\it This is not the case for
short--range interactions:}  A comparison of histograms of ground state
energies between (a) PAs and (b) RSRIM of length $L=10$
in Fig.~\ref{FigD} clearly shows the importance of long range interactions.
There is a rather clear separation of energies into `bands' with
fixed values of $Q$ for the PAs, which is almost absent in the RSRIM.
Of course, the finite width of each `band' shows that
the details of the sequence cannot be completely neglected,
although their influence on the ground state energy is rather small.

Using a Debye--H\"uckel approximation\cite{rlandau},
Wittmer {\it et al}\cite{rwittmer} have performed a systematic study
of the dependence of the free energy of neutral PAs ($Q=0$) at
high $T$  on the correlations between neighboring charges
along the chain (see Eq.~(\ref{qqcorr})).  They obtain an expression
which smoothly interpolates between the free energy densities
of a completely random sequence ($=-T\kappa^3/12\pi$, where
$\kappa^{-1}$ is the Debye screening length), and non-random
alternating sequence ($=-0.0015T\kappa^4a$). (The latter model
was also studied in Ref.\cite{rvictor}.)  These results exclude
the electrostatic self--interaction energy, which is infinite
in the continuum model used in Ref.\cite{rwittmer}.
Thus the free energy of the
alternating chain is roughly 20 times smaller than the random one.
While these results cannot be directly extended to the ground states,
we may attempt to obtain crude estimates by setting $\kappa^{-1}=a$
and $T=q_0^2/a$. However, our results indicate that the ground state
energy of alternating polymers ($=\epsilon_{c0}$) is only smaller
by about 16\% than the mean condensation energy of unrestricted sequences.
Such inconsistency is partially explained
by the fact that the alternating PA has negative mean
electrostatic energy (approximately $0.6q_0^2/a$ per atom)
even at $T=\infty$, while such an energy for a completely
random PA (averaged over all quenches) vanishes. Thus,
only about 1/4 of the ground state energy of an
alternating PA is its condensation energy. (This part of the energy
explicitly depends on the discreteness of the chain and is not accounted
for in Ref.\cite{rwittmer}.) This argument
brings the approximate conclusions based on Ref.\cite{rwittmer}
in better qualitative agreement with our exact enumeration
results.

The dependence of the quench--averaged ground state
energies on the length of the chain is depicted in Fig.~\ref{FigE}.
A restricted average is performed at each value of $Q$ (indicated
next to each line). The scaling of the axes is motivated by the
re-casting of  Eq.~(\ref{edefenerg}) in the form
\begin{equation}\label{EQNscale}
{E\over N}-{A\over N^{1/3}}=-\epsilon_c+{bQ^2\over aN^{4/3}}\ ,
\end{equation}
where $A\equiv\gamma S/N^{2/3}=p\gamma a^2$, with a prefactor
$p$ depending on the average shape. The small number of data
points makes an accurate determination of $A$ (and hence the
surface tension) rather difficult.  The value used in Fig.~\ref{FigE}
is $A=0.6q_0^2/a$, for which the curves with different  $Q$ extrapolate
to approximately the same value, giving a condensation energy of
$\epsilon_c=(0.75\pm0.01)q_0^2/a$. For this choice of $A$ the slopes
of the curves with $Q=1,2,3$ approximately scale as $Q^2$.
The condensation energy $\epsilon_c$ is surprisingly close to that
of a regular crystal ($\epsilon_{c0}=0.8738q_0^2/a$), despite the fact
that in a random chain on  average one neighbor (along the chain)
has the ``wrong'' sign (compared to the alternating arrangement), costing an
energy of the order of $\epsilon_{c0}$. This again confirms our contention
that the ground state energy is determined by very extended
neighborhoods of each particle. If the ground state configuration has
approximately cubic or spherical shape, then $p\approx 5$, while for
the slightly elongated objects that we obtain, $p$ can be somewhat larger
($\approx8$). Therefore, we estimate $\gamma=(0.09\pm0.03)q_0^2/a$.
The error bars indicate our uncertainty in the values of $p$ and $A$,
and disregard possible systematic errors in attempts to evaluate surface
tension from such small clusters.
Using these numbers we estimate that the Rayleigh charge of the model PA
is approximately the same as $Q_c$, since
\begin{equation}
Q^2_R=12 \gamma V\approx 12 \times{0.09q_0^2\over a}\times aN^3\approx q_0^2N
=Q_c^2\ .
\end{equation}
(The relation $V=a^3N$ assumes PAs of maximum possible density.)
{}From Fig.~\ref{FigE} it is not clear that the (charge unconstrained) average
energies (indicated by the $\times$ symbols) of all quenches, also
extrapolate to the same condensation energy of $\epsilon_c$.
This apparent inconsistency can be understood by noting that since
the quench averaged $Q^2$ is equal to $q_0^2N$, the last term in
Eq.~(\ref{EQNscale}) scales as $N^{-1/3}=(N^{-4/3})^{1/4}$. Thus, the
linear approach (in the variables used in Fig.~\ref{FigE}) to asymptotic
value (as $1/N^{4/3}\rightarrow 0$) is replaced by a very small
power law. Such a slow decay cannot be detected for the small values of
$N$ used in our enumeration  study.

Since our model is defined on a discrete lattice, the allowed energies
are discrete. However, as the length of the chain increases the
separations between the states are reduced. The density of states becomes
quasi--continuous and can be described by a function $n(E)$.
Fig.~\ref{FigF} depicts $s(E)\equiv\ln\overline{n(E)}/N$, where the overline
denotes averaging over all quenches with a fixed $Q$.
(Note that this quantity is {\it not} the quench--averaged free energy as
the average is performed on $n$ rather than on $\ln n$.) Not surprisingly, the
densities of states for different $Q$s are shifted with respect to each other.
For every quench the density of states is very high near the middle of the band
and decreases towards the edges.

We find that almost all PAs have a unique ground
state (up to trivial symmetry transformations). This is not the case for
short range interactions\cite{groundstate} and may be an important clue
to the problem of protein folding. (For ease of calculation, most studies
of similar random copolymers have focused on short-range interactions,
and typically find highly degenerate ground states.)
Furthermore, the gaps to the second lowest energy states typically remain
of order of $0.1q_0^2/a$ (up to the studied size of $L=12$), while most
interstate separations decrease with $L$.
In the $L\to\infty$ limit, the density of lowest energy excitations of our
model
PAs appears to decay faster than a power law. (Of course our lattice model
does not include any vibrational modes.)
This decay manifests itself in a vanishing heat capacity in the $T\rightarrow0$
limit, as depicted in Fig.~\ref{FigG}. The solid lines represent the
quench averaged heat capacities per degree of freedom $c$, of PAs with
$Q=0$ at low temperatures. (Since the energy fluctuations of a polymer depend
only on changes of its shape, and are independent of its overall position
and orientation, we assumed that an $N$--atom
PA has $3N-5$  degrees of freedom, where $-5$ represents subtraction
of translational and rotational degrees of freedom. Such a choice decreases
the bias in the $N$--dependence of $c$ which would appear for very small
values of $N$.)
The vanishing heat capacity was {\it not} observed in MC
studies\cite{KKshort,KKlong},
where poor equilibration at low $T$ hinders measurement of $c$.

It is instructive to compare and contrast the behavior of random PAs with
vanishing excess charge to that of an ordered alternating sequence;
the latter is a highly atypical member of the
ensemble with $Q=0$. Numerical investigations of alternating charge
sequences by Victor and Imbert\cite{rvictor} show that such polymers
undergo a collapse
transition, similar to SAWs with {\it short range} attractive interactions.
This is because  the exact compensation in the charges of any pair of
neighboring monomers leads to large scale properties determined by
dipole--dipole (and faster decaying) interactions. Thus Coulomb interactions
are irrelevant in the high temperature phase of the alternating chain that
consequently behaves as a SAW. By contrast, even though we consider
a sub--ensemble of quenches with $Q=0$,
the charge fluctuations cannot be neglected in random PAs and control
the long distance behavior of the chain. Such PAs are compact at {\it any}
temperature.

The attractive dipole--dipole interactions eventually cause the collapse
of the alternating charge sequence to a compact state at
temperatures below a $\theta$--point.
Of course, the ground state of such a chain is the ordered {\tt NaCl} crystal
discussed earlier. However, it is not clear if the state of the chain
immediately below
the $\theta$ temperature is the ordered crystal. Another possibility
is that the initial collapse is into a ``molten globular" (liquid like) state
\cite{rstein},
which then crystallizes at a lower temperature. We singled out the alternating
PAs in our complete ensemble of quenches; the dashed lines in Fig.~\ref{FigG}
depict the heat capacity of this sequence. The presence of a phase
transition manifests itself in the peak in $c$ at $T\approx0.14q_0^2/a$ (for
$L=11$)
which grows (and slightly shifts towards higher temperatures)
as $L$ increases.
Fig.~\ref{FigG} shows that the average heat capacity of random PAs with $Q=0$
also has a peak at $T\approx0.17q_0^2/a$  (for $L=11$).
As the high temperature phase is no longer swollen (for $Q=0$), there are
again two possible interpretations of this heat capacity peak. One is that
it represents a crossover remnant of the $\theta$ transition, with an increase
in the density of the compact polymer. Indeed, the peak  is lower and
broader than that of alternating chains. Another possibility is that there
is a ``glass" transition in which the ``molten globule" freezes into its
`ground state'. The proximity of  the peak temperature to the energy gap for
the first excited state supports the latter conclusion.

No corresponding anomaly was observed in the MC
simulations\cite{KKshort,KKlong}.
Since finite size effects are extremely important in such small systems, the
heat
capacity peak should  be regarded only as a suggestion for the presence of a
``$\theta_E$--transition". As indicated by the dashed line in Fig.~\ref{FigA},
the
location of such a transition may depend on $Q$, disappearing at
$Q\approx q_0\sqrt{N}$, consistent with other features of the phase diagram.
This behavior is analogous to that of the $\theta$--point in the
RSRIM\cite{rsrim}, although in that case the limiting charge scales linearly
with $N$. Additional, studies are needed to establish the
$\theta_E$--transition.

\section{Polymer Shapes}\label{secshape}

The contour plots in Fig.~\ref{FigH} depict the number of states as a function
of
both $R_g^2$ and $E$, for three sequences of $L=10$ with charges $Q=1$
(a), 5 (b), and 11 (c). At high temperatures the typical configurations
correspond
to the highest densities.  In all three cases these configurations
are located in the middle of the diagram, and behave essentially as SAWs.
On lowering temperature the polymer seeks out states of lowest energy
which are very different in the three cases. The approximately neutral
chain of Fig.~\ref{FigH}a assumes a very compact shape
represented by the lower--left corner of the contours.
The presence of a specific heat peak is consistent with
the shape of this contour plot. While $R_g^2$ increases monotonically
with $T$, the chains are too short to permit a quantitative test for the
presence of  a $\theta_E$--point from the scaling of $R_g^2$.
The lowest energy contour of the chain with $Q=5$ (Fig.~\ref{FigH}b)
is almost horizontal. Hence, upon lowering temperature the chain
will not collapse, maintaining an extended shape. Thus a putative
transition must  disappear for larger $Q$.
Finally, the fully charged polymer in Fig.~\ref{FigH}c expands from a SAW
to the completely stretched configurations represented by the lower--right
corner of the contour plot.

The low temperature results from MC simulations suggest that
$R_g^2$ of a PA strongly depends on its charge, crossing over
from compact configurations at small $Q$ to extended states for
larger $Q$. This is qualitatively supported by the ground state shapes
in Fig.~\ref{FigB}, and will be more quantitatively examined here.
Fig.~\ref{FigI}a depicts the $L$--dependence of $R_g^2$ for
several choices of $Q$. The vertical axis is scaled so that
compact, i.e. fixed density, structures are represented by
horizontal lines. Since $Q$ is fixed, the influence of the excess charge
diminishes as the length of the polymer is increased, and thus all curves
must asymptotically converge to the same horizontal line. There is
some indication of this in Fig.~\ref{FigI}a, although the crossover is
rather delayed for larger values of $Q$. Since the unrestricted ensemble
(solid circles) includes a large range of $Q$s, it is not surprising that the
corresponding averages are not compact. The chains are too
short to extract a meaningful value for the exponent $\nu$.
Nevertheless,  the effective slope of $\nu_{\rm eff}\approx {1/2}$, strongly
suggests that the average over an unrestricted ensemble is not compact.
By comparison, the corresponding results for the RSRIM in Fig.~\ref{FigI}b
clearly indicate that the averages both at fixed and varying $Q$ have
similar fixed density ground states.

Since the quench--averaged $R_g^2$ of the unrestricted
ensemble scales differently from the sub--ensembles of fixed $Q$,
the former set must contain a non--negligible portion of non--compact
configurations for every $L$ .  It is natural to assume that
the borderline between compact and stretched states is controlled
by $Q_c=q_0\sqrt{N}$.  In previous work\cite{KKshort,KKlong} we argued that PAs
undergo a transition to an expanded state when $Q$ exceeds $Q_R$
($\propto Q_c$): the transition is more pronounced for larger
$L$ and lower $T$. In the MC simulations\cite{KKshort,KKlong}  we were
able to use long PAs, but were restricted to  finite, albeit small,
temperatures
which slightly smeared the transition in $R_g$ with increasing $Q$.
In this study we know the exact ground states but are limited to small
$L$s where the difference between $R_g$ of compact and stretched states
is less visible. The sum of all eigenvalues of the shape tensor,
$R_g^2$ is  somewhat insensitive to an expansion since the increase in the
largest eigenvalue is partially compensated by the decrease of the other
two eigenvalues. A clearer view is provided by the ratios of the mean
eigenvalues
of the  shape tensor as depicted in Fig.~\ref{FigJ}. These ratios for different
$L$s can be collapsed after scaling the charges by $Q_c$, consistent with
the MC simulations.

Fig.~\ref{FigK} depicts (on a logarithmic scale) the distribution
of values of $R_g^2$ in the ground states of all quenches for
$L=13$. The distribution is peaked near the smallest possible
value of $R_g^2$, but has a broad (possibly power law) tail.
If the tail falls off sufficiently slowly, it will determine the asymptotic
value of the exponent $\nu$:  As $L$ increases the
very large values of $R_g^2$ of the (minority) stretched
configurations will eventually dominate the total average.
We thus expect $\nu_{\rm eff}$ to increase with $L$, and the value
of $\nu_{\rm eff}$ extracted from the slope of the solid
line on Fig.~\ref{FigI}a, probably underestimates the true
asymptotic value.  To get further insight into the behavior for larger $L$,
we performed separate averages for the 80\% of configurations which
have the smallest $R_g^2$, and for the remaining top 20\%. These averages
are depicted in Fig.~\ref{FigL}. The vertical axis is again scaled so that
compact structures are represented by horizontal lines. The bottom 80\%
indeed scale as compact chains while the top 20\%, which stand for the
tail of the distribution, have radii that grow with $L$ with an effective
exponent
of $\nu_{\rm eff}\approx {2/3}$. We thus conclude that the
$R_g^2$ of the unrestricted ensemble increases with
$L$ at least as fast as a SAW.

\section{Annealed Polyampholytes}\label{secanneal}

As a byproduct of our study, since we have access to the complete
set of quenches, we can find which particular sequence, restricted
only by its net charge, has the lowest energy.  As this is the sequence
that is selected in a model in which the charges are  free to change
positions along the chain, we shall refer to the results as describing
the ground states of {\it annealed PAs}. For long chains, neither the
sequence, nor its spatial conformation, need to be unique.
However, for the sizes considered here, we always found a
single ground state, several of which are shown in Fig.~\ref{FigM}
for $L=13$ and different values of $Q$. It appears that the optimal
configurations correspond to a uniform distribution of excess charge
along the backbone. In particular, for small $Q$ the preferred
arrangement is the alternating sequence which then folds into a
{\tt NaCl} structure.

Previously\cite{KKshort,KKlong} we suggested that
annealed PAs expel their excess charge  (provided
$Q>q_0N^{1/3}$) into highly charged ``fingers''. As a result of
such ``charge expulsion'' the spanning length of annealed PAs
should increase dramatically ($\sim Q$). However, since
most of the mass  remains in a compact globule, $R_g^2$
is not substantially modified (as long as $Q<Q_c\sim q_0N^{1/2}$).
The chains used in our study are too short to exhibit an increased
spanning length with no change in $R_g$. Moreover, the effects of
lattice discreteness are much more pronounced for annealed PAs
where ground states correspond to a single sequence. In the
quenched case, averaging over all sequences partially smoothens
out lattice effects. As partial evidence we note that plots for the charge
dependence of ratios of eigenvalues of the shape tensor (analogous
to Fig.~\ref{FigJ}) exhibit better collapse  with the variable $Q/N^{1/3}$
than with $Q/N^{1/2}$. However, given the scatter of the few data points,
the evidence for the appearance of ``fingers'' is not really any more
convincing than any conclusions drawn from inspection
of the ground states in Fig.~\ref{FigM}.

As noted earlier, we expect the ground state of a sufficiently long annealed
PA with fixed $Q$ to be the {\tt NaCl} structure. To test the approach
to this limit, in Fig.~\ref{FigN} we plot the energies per atom of the ground
states. As in the case of quenched PAs (Fig.~\ref{FigE}), we check for finite
size corrections proportional to the surface area. (Unlike the case of quenched
PAs, each point in this figure represents a {\it single} configuration.) Here
we used a value of  $A=0.35q_0^2/a$ although the results are rather insensitive
to this choice, and we estimate the accuracy of this quantity as
$\pm0.10q_0^2/a$. The point of intersection with the $1/N=0$
axis is close to the known value of the $\epsilon_{c0}$.
Furthermore, $A=(0.35\pm0.10)q_0^2/a$ corresponds to a surface tension of
$0.05\pm0.03q_0^2/a^3$ which is also consistent with the known value
of $\gamma_0$. These consistency checks add further confidence to the values
of $\epsilon_c$ and $\gamma$ deduced for quenched PAs.

\acknowledgments
This work was supported by the US--Israel BSF grant
No. 92--00026, by the NSF through grants No. DMR--94--00334 (at MIT's
CMSE), DMR 91--15491 (at Harvard), and the PYI program (MK).

\appendix

\section{Rayleigh Instability in Arbitrary
Space Dimensions} \label{secrayleigh}

In this Appendix we discuss instabilities of charged
$d$--dimensional drops. A detailed discussion of the
three--dimensional case can be found in Appendices B and
C of Ref.\cite{KKlong}, which also provides other
references to the subject.
The energy of a charged {\it conducting} (hyper)sphere of radius
$R$ with charge $Q$ is given by
\begin{equation}\label{eEQ}
E=\gamma S_dR^{d-1}+{Q^2\over 2R^{d-2}}\ ,\quad (d>2)
\end{equation}
where the first term is the surface energy  ($\gamma$ is the surface tension,
and $S_d$ denotes the $d$--dimensional solid angle), while the second term
is the electrostatic energy. (We have used units such that, in $d$ dimensions,
the electrostatic potential at a distance $r$ from a charge $q$ is
$q/r^{d-2}$; and $q\ln r$ in $d=2$.)
For small $Q$, the sphere is stable with respect
to infinitesimal shape perturbations. However, when the electrostatic and
surface energies are comparable, the drop becomes unstable. To
explore this instability we differentiate  Eq.~(\ref{eEQ}) with respect  to
$R$ to find the pressure difference between the interior and the exterior
of the drop as
\begin{equation}
\Delta p={(d-1)\gamma\over  R}-
{(d-2)Q^2\over 2 S_d R^{2d-2}}\ .
\end{equation}
The pressure difference vanishes when $Q$ equals
the {\it Rayleigh charge} $Q_R$, where
\begin{equation}\label{eQR}
Q_R^2\equiv {2(d-1)\over d-2} S_d \gamma R^{2d-3}\ .
\end{equation}
For $Q\geq Q_R$ a (hyper)spherical shape is unstable
to small perturbations; initially the drop  becomes
distorted and subsequently it disintegrates. Note that
$Q_R^2\sim R^{2d-3}\sim V^{2-3/d}$, where $V$ is the
volume of the drop. When applied to PAs, up to a
dimensionless prefactor,
\begin{equation}\label{edefQRind}
Q_R^2\approx q_0^2N^{2-3/d} \ .
\end{equation}

We can regard the first term in Eq.~(\ref{eEQ}) as setting the overall
energy scale, while the shape of the drop is determined by the
dimensionless ratio
\begin{equation}\label{ealphadef}
\alpha\equiv Q^2/Q^2_R \ .
\end{equation}
While from the above argument we conclude that  the spherical shape is
(locally) unstable for $\alpha\geq1$, even for $\alpha<1$, the energy of
the drop can be lowered by splitting into smaller droplets. In particular,
we may split away from the original drop a large number $n$, of small droplets
of
radius $R/n^x$ and charge $Q/n$, and remove them to infinity. It can be
directly
verified that for $1/(d-1)<x<1/(d-2)$, the total electrostatic energy, total
surface
area of the small droplets, as well as their total volume, vanishes in the
$n\rightarrow\infty$ limit. Thus the energy of any charged conducting drop can
be lowered to that of an uncharged drop by expelling a large number of
``dust particles'' which carry away the entire charge. (Of course this argument
neglects the finite size of any particles making up the drop!)

The globular phase of a quenched random PA is better
represented by a drop of immobile charges. Therefore,
we next consider a drop in which the charges are {\it uniformly}
distributed over the volume. The sum of the surface and
electrostatic energies  is now given by
\begin{equation}
E_1=\gamma S_dR^{d-1}+{d\over d+2}{Q^2\over R^{d-2}}\ .
\end{equation}
For sufficiently large $Q$, the drop can lower its energy by splitting into
two droplets of equal size. This will occur when $\alpha$ exceeds a
critical value of
\begin{equation}
\alpha_d={2^{1/d}-1 \over 1-2^{-2/d}}
{(d^2-4)\over 2d(d-1)}\ ,
\end{equation}
which is equal to 0, 0.293, 0.323, 0.322,
and $1/4$, for $d=2$, 3, 4, 5, and $\infty$, respectively.
As the value of $\alpha$ increases further, the drop splits into a larger
number of droplets. By examining the energy of a system of  $n$ equal
spherical droplets, we find that the optimal number is proportional to
$\alpha^{d/3}$. If the typical $Q^2$ is proportional to $N$ (as happens
in unrestricted PAs), while $Q_R^2$ is given by Eq.~(\ref{edefQRind}),
the number of droplets scales as $N^{1-d/3}$. Thus $d=3$ is a special
dimension, above which a typical PA prefers to stay in a single globule.

\begin{figure}
\caption{Qualitative phase diagram of a PA as a function
of temperature $T$ and its excess charge $Q$. See the text
for details.}
\label{FigA}
\end{figure}
\begin{figure}
\caption{Spatial conformations of ground states
for a sample of $L=12$ PAs, for values of $Q/q_0$ equal to (a) 1, (b) 3,
(c) 7, and (d) 9. Dark and bright shades indicate opposite
charges.}
\label{FigB}
\end{figure}
\begin{figure}
\caption{Energies of the ground states\protect\cite{units} of
all distinct (i.e. unrelated by symmetry
transformations) quenches (arbitrarily numbered from 1
to 2080) of PAs with $L=12$.}
\label{FigC}
\end{figure}
\begin{figure}
\caption{Histograms of the numbers of ground states versus their
energies\protect\cite{units}  for chains of length $L=10$ with
charges $Q=1$ (dotted line), $Q=3$ (solid line), $Q=5$
(dashed line), and $Q=7$ (dot--dashed line) for (a) PAs and
(b) polymers with random short range interactions. The size
of the energy bins for these histograms is $\Delta E=1$. }
\label{FigD}
\end{figure}
\begin{figure}
\caption{Ground state energy (per particle)\protect\cite{units}
as a function of the number of particles. Each full circle
represents a quench average at the values of $Q$ denoted by the
numbers near the solid lines. ($Q=N$ corresponds to
a fully charged polyelectrolyte.) The $\times$ symbols
denote the average over all quenches, unrestricted in charge.}
\label{FigE}
\end{figure}
\begin{figure}
\caption{Logarithms of quench--averaged densities of states
per atom (see text) for chains of $L=13$  with the
excess charge $Q$ set to 1, 3, 5, $\cdots$, 13 (from left to right).}
\label{FigF}
\end{figure}
\begin{figure}
\caption{The quench averaged heat capacity per degree of freedom
for a random PA with $Q=0$ (solid lines) and $L=3,$ 5, 7, 11
(from bottom to top); and for alternating--charge sequences
(dashed lines) of similar lengths.}
\label{FigG}
\end{figure}
\begin{figure}
\caption{Contour plots of the number of levels as a function of $R_g^2$
(in lattice constants) and $E$ (in units of  $q_0^2/a$). The size of each bin
is 0.1 in the $E$--direction, and $0.25$ in the $R_g^2$
direction. The contours represent continuous interpolations at
levels 0.5, 33, 129, 513, 2049, and 8193. Each plot represents
a single chain with $L=10$ and excess charge $Q$ of (a) 1, (b) 5, and
(c) 11.}
\label{FigH}
\end{figure}
\begin{figure}
\caption{Quench--averaged scaled $R_g^2$ as a function of the length of
the polymer for (a) PAs with electrostatic interactions, and (b)
random polymers with local interactions.
Open circles and dashed lines represent (from bottom to top)
the restricted averages with $Q=0$, 1, 2, 3. Full circles and the solid line
represent unrestricted averages. }
\label{FigI}
\end{figure}
\begin{figure}
\caption{Ratios between  the smallest and largest eigenvalues
of the shape tensor (dashed lines), and between the intermediate
and largest eigenvalues (solid lines), for $L=8$, 10, and 12
(full triangles, squares, and circles respectively) as a function of the scaled
charge $Q/Q_c$. The eigenvalues are calculated from
quench averages at fixed $Q$.}
\label{FigJ}
\end{figure}
\begin{figure}
\caption{Probability distribution for $R_g^2$ of
ground states in the unrestricted ensemble of quenches
for $L=12$. $R_g^2$ is scaled by  its minimal possible value,
while the scale of the probability is arbitrary.}
\label{FigK}
\end{figure}
\begin{figure}
\caption{Scaled partial averages of $R_g^2$ of the ground states
for the 20\% of quenches with largest $R_g$s (top curve), and the
remaining quenches (bottom curve) as a function of the length
of the PA.}
\label{FigL}
\end{figure}
\begin{figure}
\caption{Spatial conformations of the ground states
of annealed PAs with $L=12$, for values of $Q/q_0$ equal to (a) 1, (b) 3,
(c) 5, and  (d) 7. Dark and bright shades indicate opposite
charges.}
\label{FigM}
\end{figure}
\begin{figure}
\caption{Ground state energy per atom \protect\cite{units}
for annealed PAs
as a function of polymer length. Each point represents a single
configuration for $Q=0$, 1, 2, and 3, denoted
by open and full circles, and open and full triangles, respectively.}
\label{FigN}
\end{figure}

\begin{references}
\bibitem{tcompare}The Coulomb potential used in
Refs.\protect\cite{KKshort,KKlong}  was slightly distorted at short distances,
with different short range cutoff and intermonomer distances. Therefore,
there is no exact equivalence between $T$ of this work and that in
Refs.\protect\cite{KKshort,KKlong}. An approximate correspondence is
obtained by multiplying the $T$s in Refs.\protect\cite{KKshort,KKlong} by a
factor of 2 to 3.
\bibitem{units}In all figures in this paper lengths are
measured in units of $a$, charges in units of $q_0$,
temperatures and energies in units of $q_0^2/a$. For the RSRIM,
charges are dimensionless, and energies are instead measured in
units of $v_0$.
\end{references}
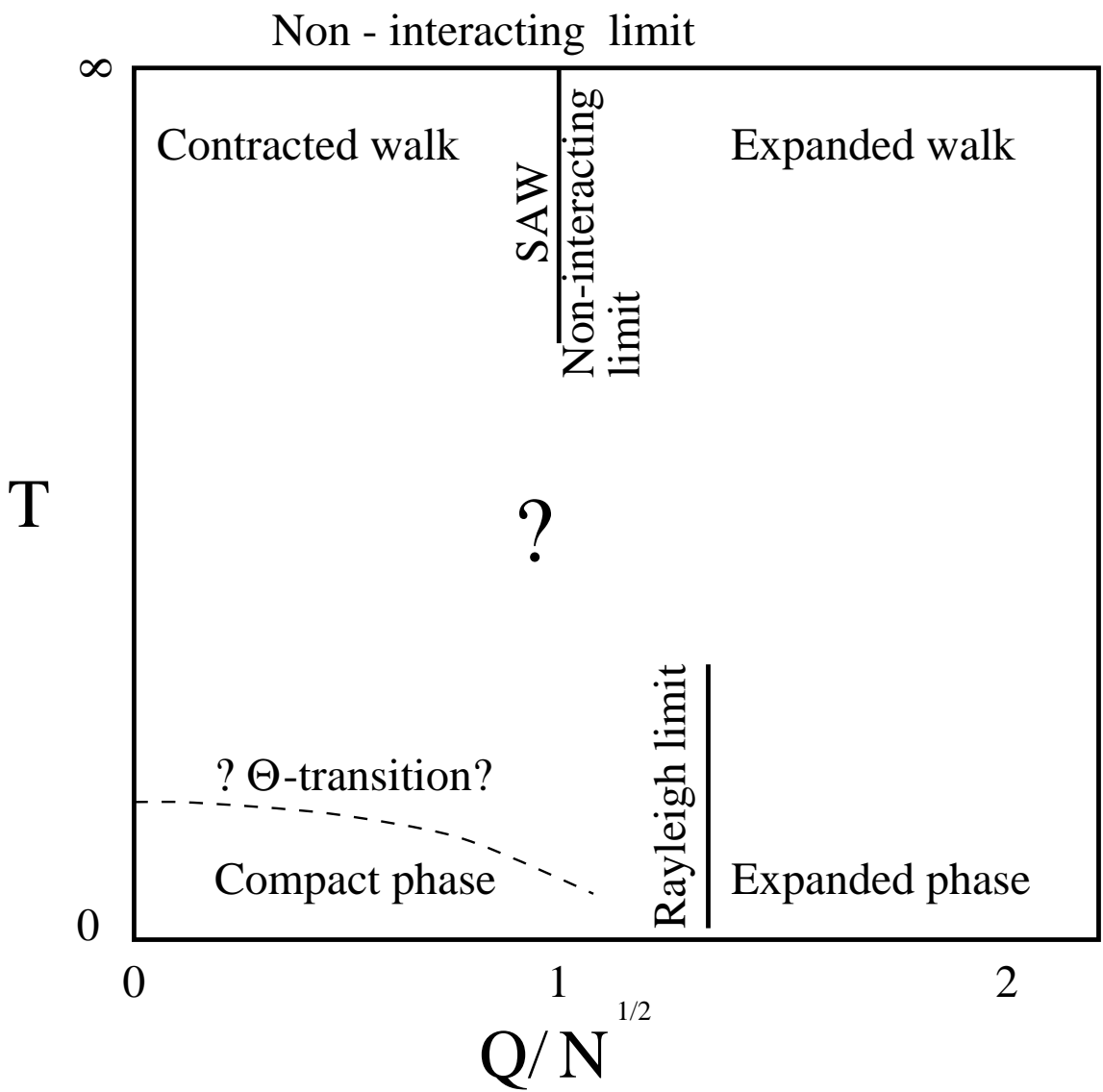
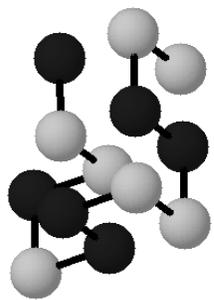
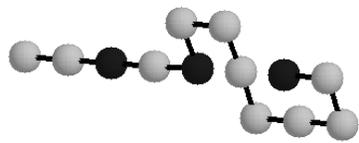
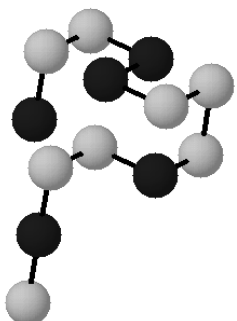
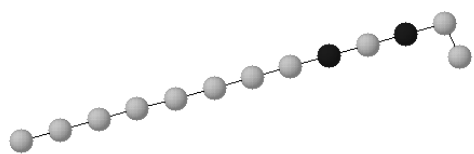
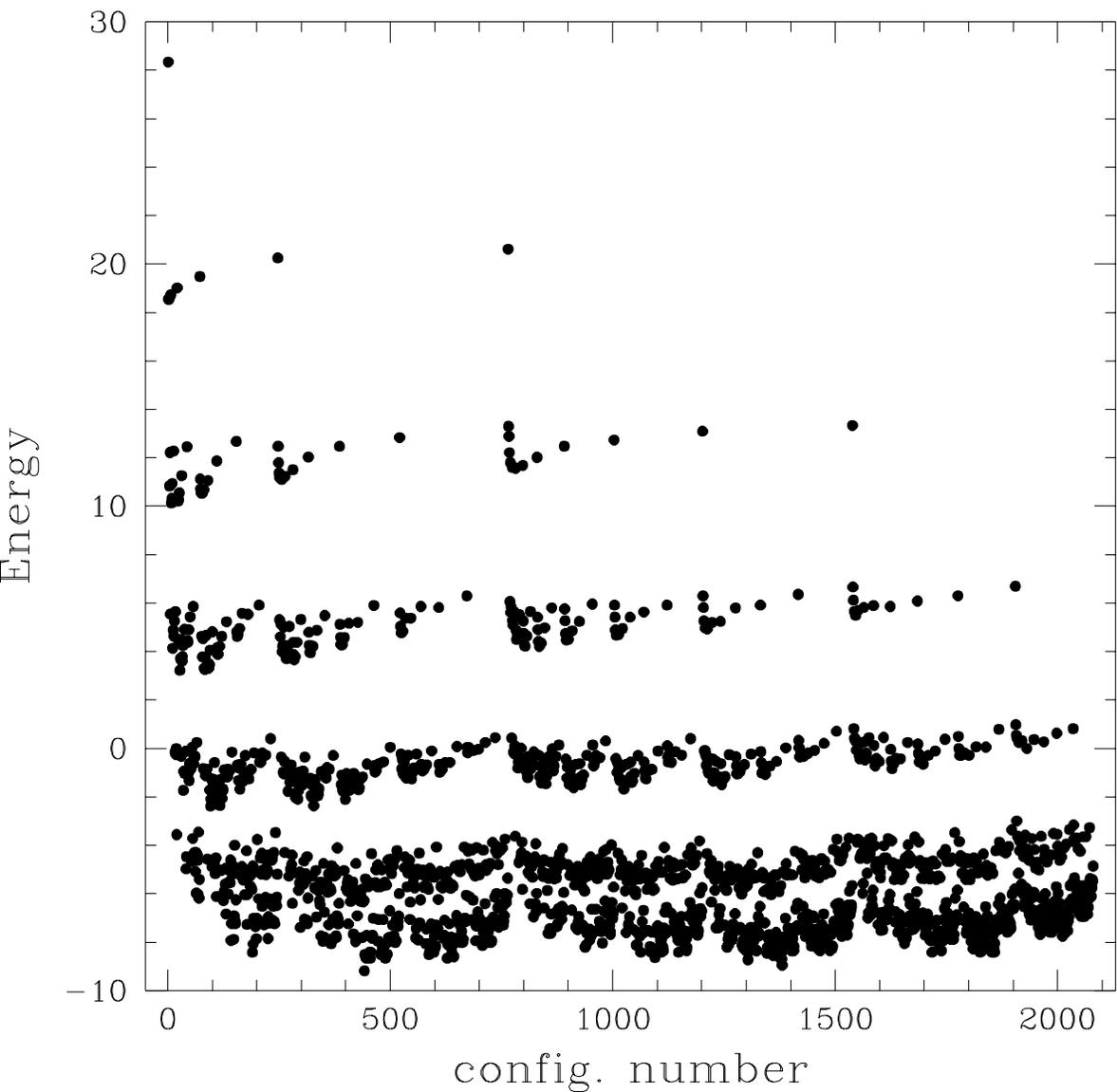
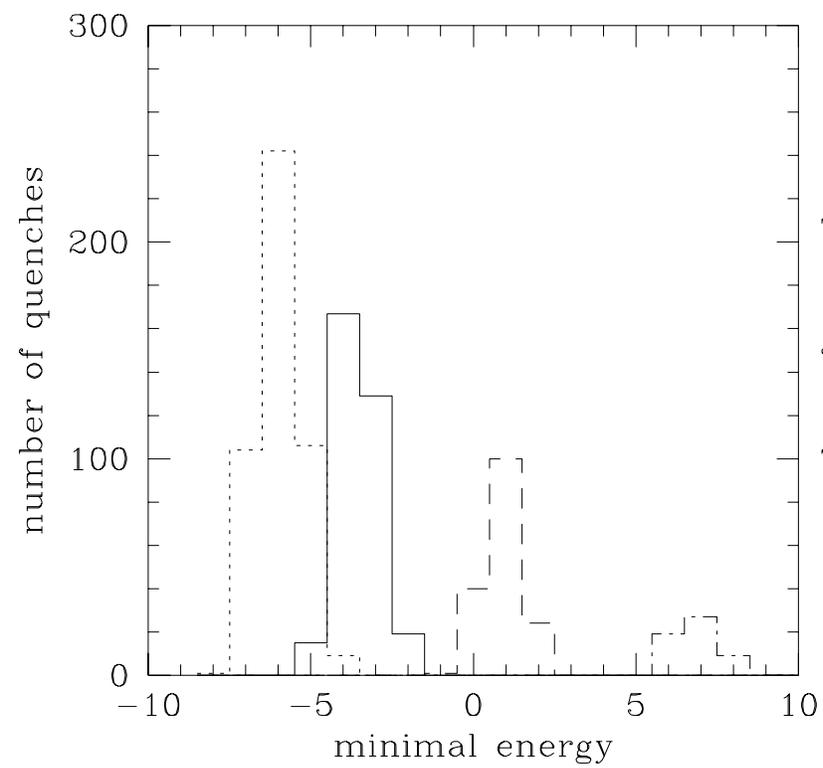
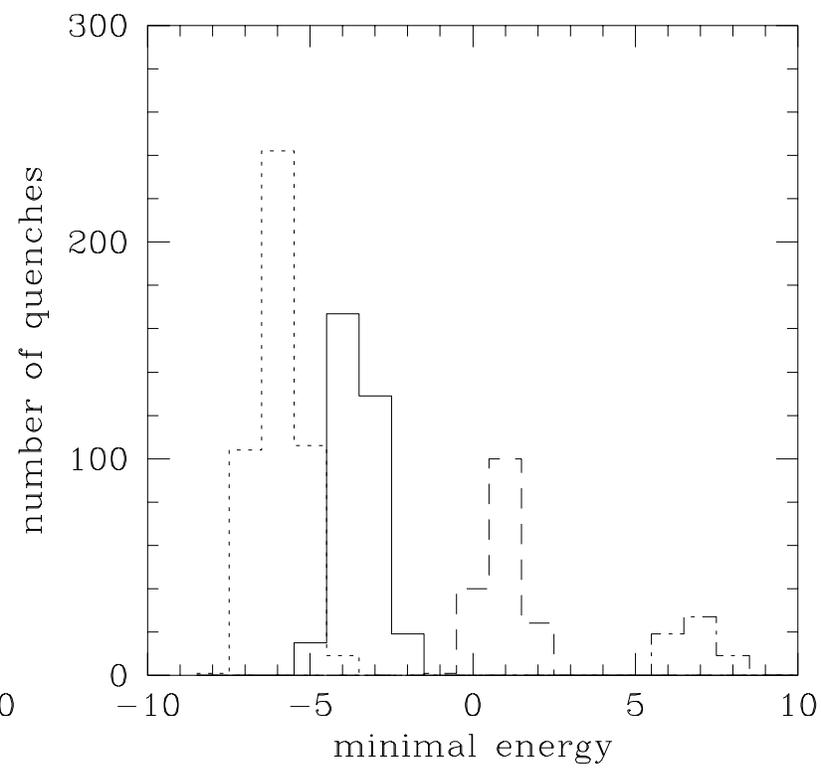
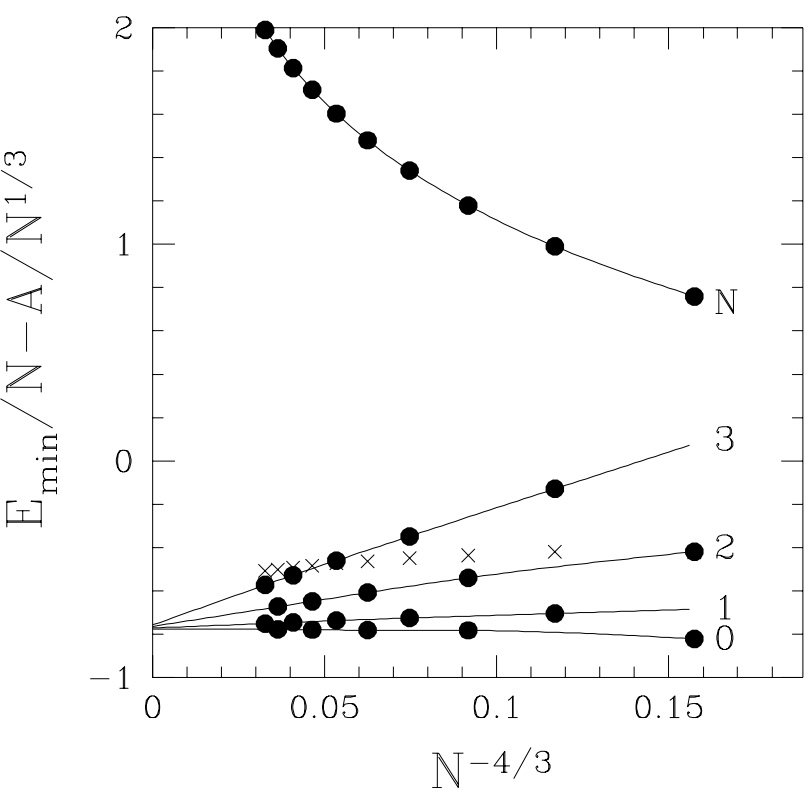
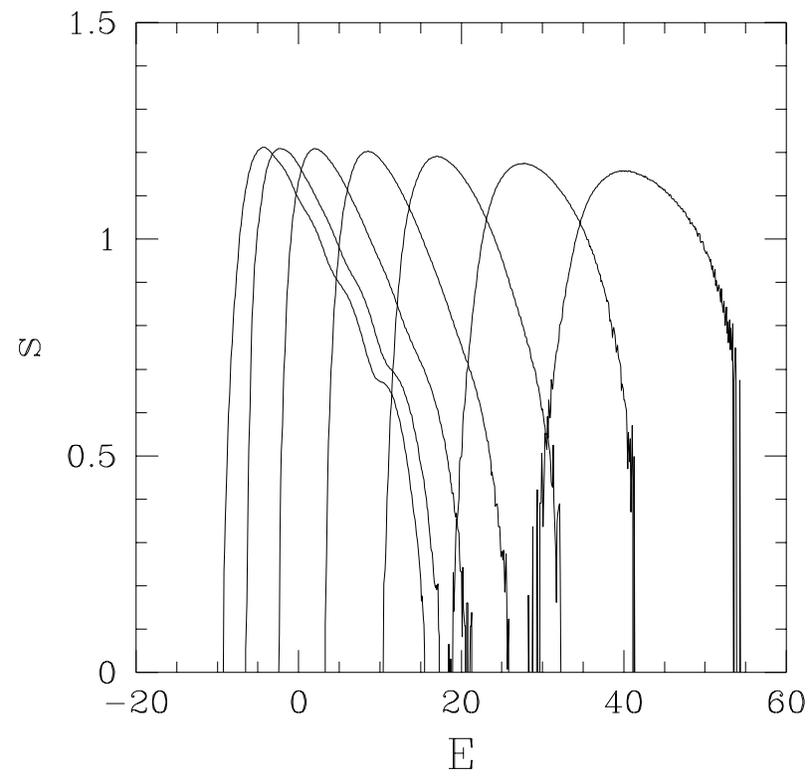
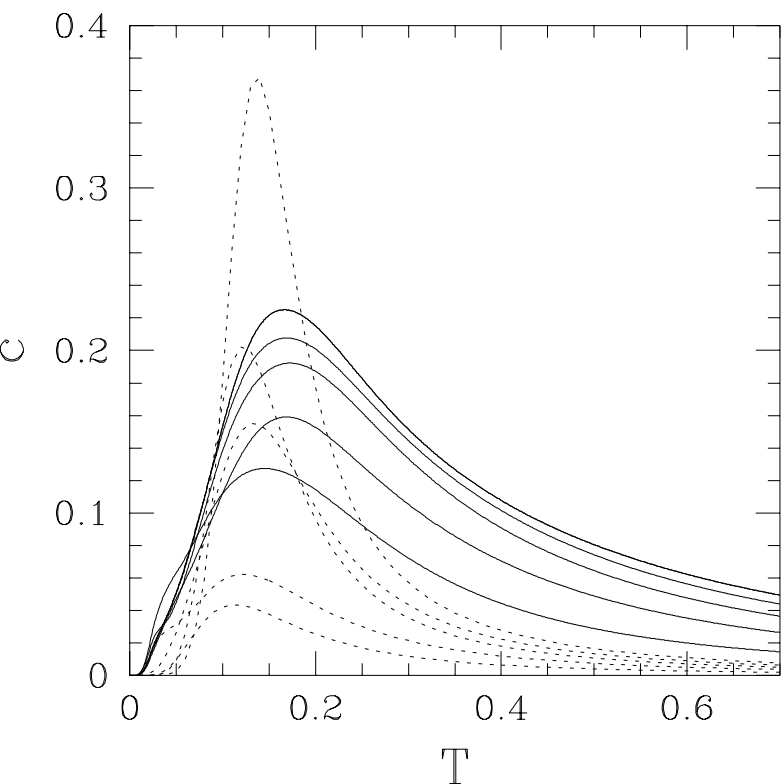
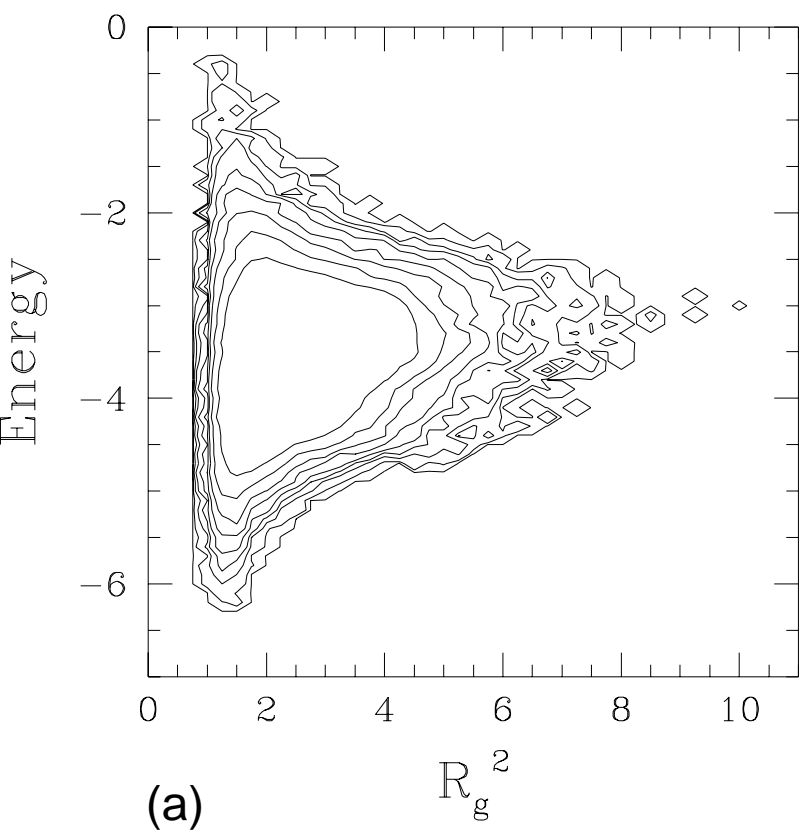
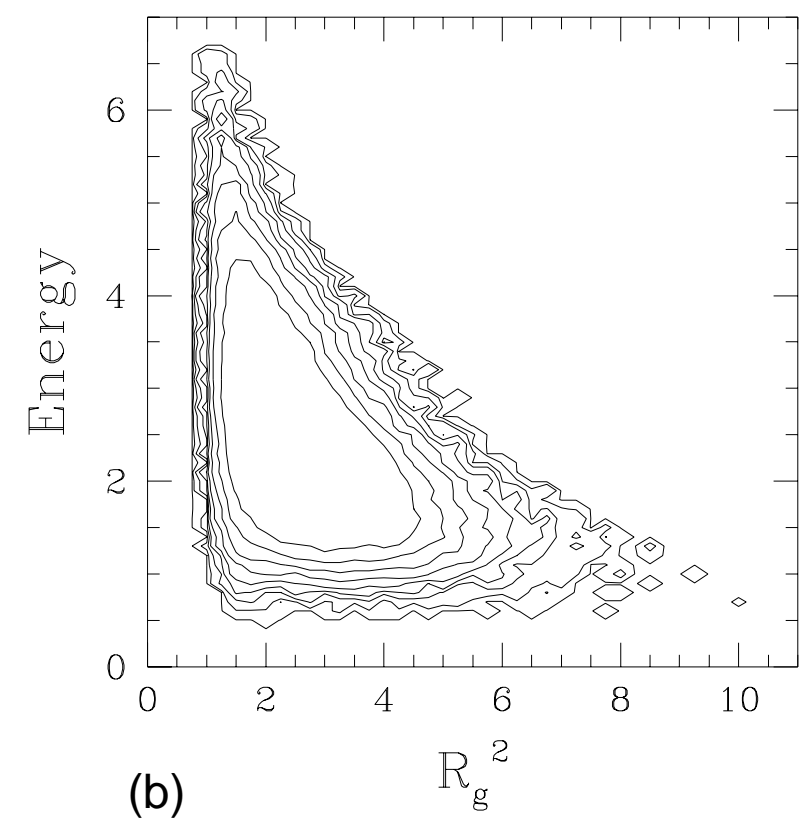
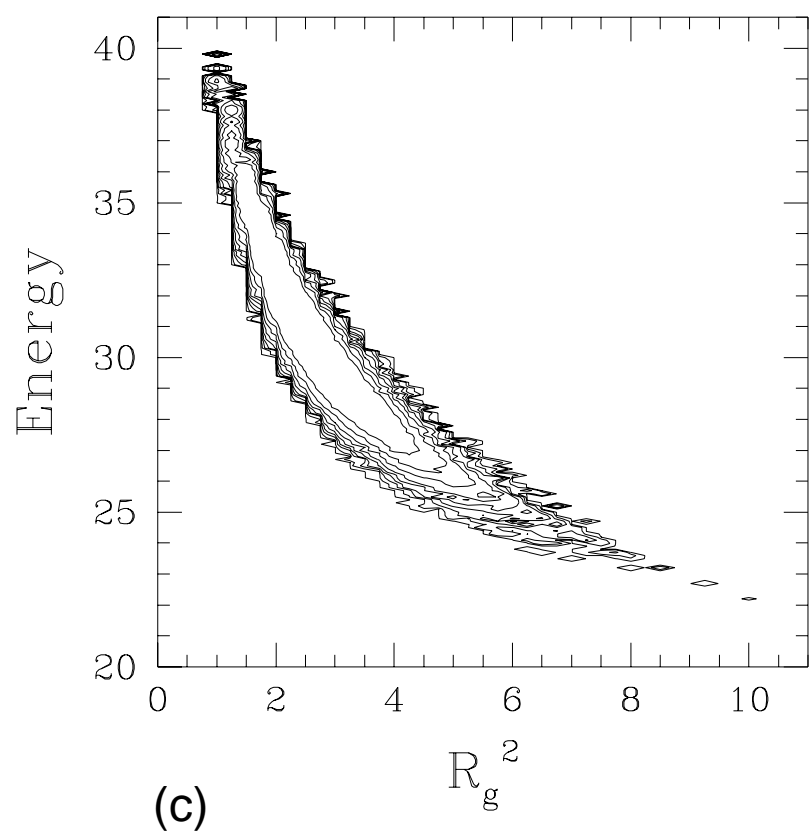
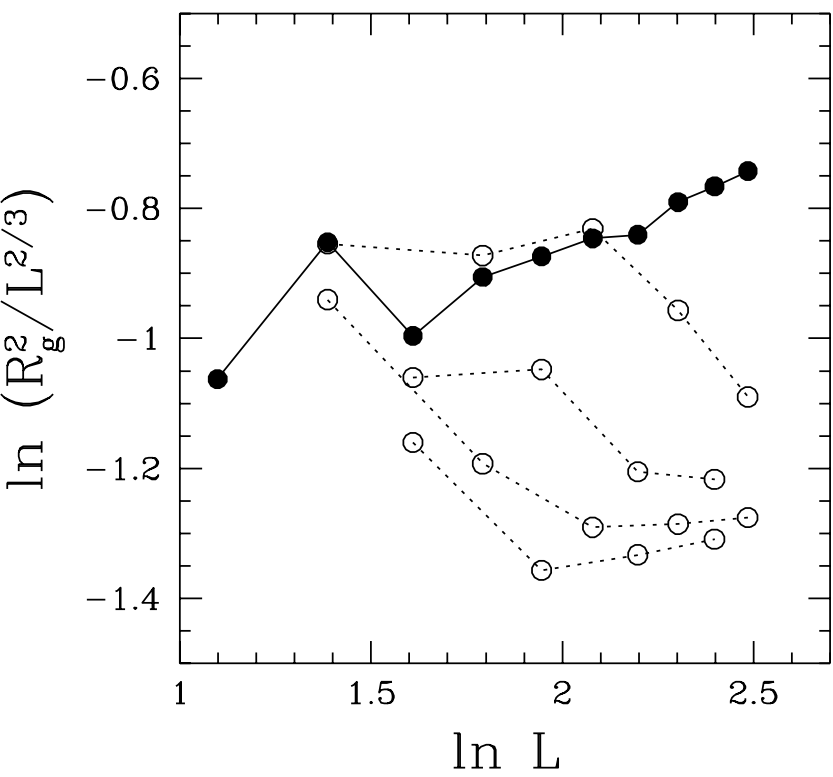
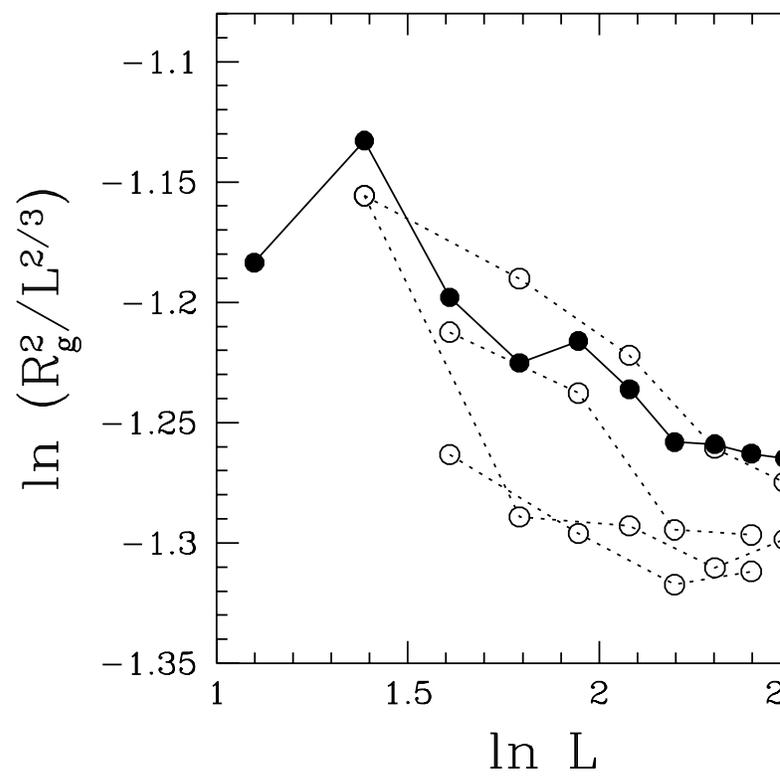
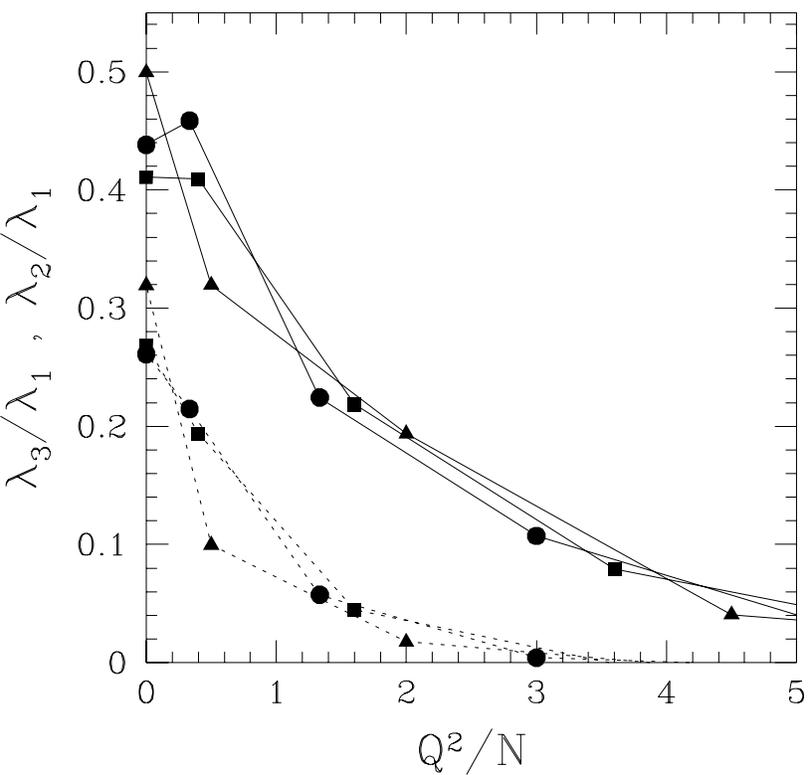
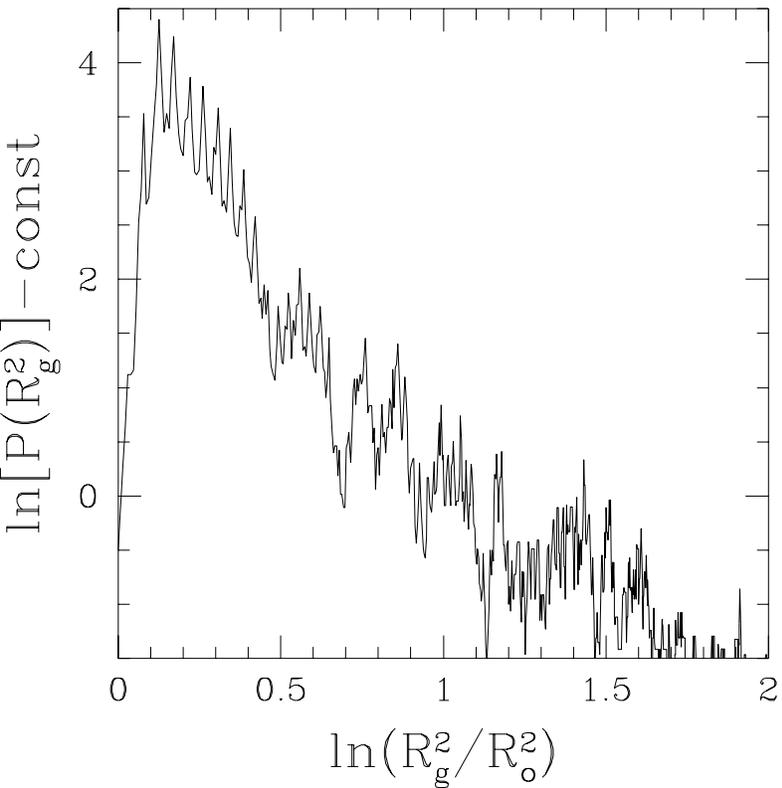
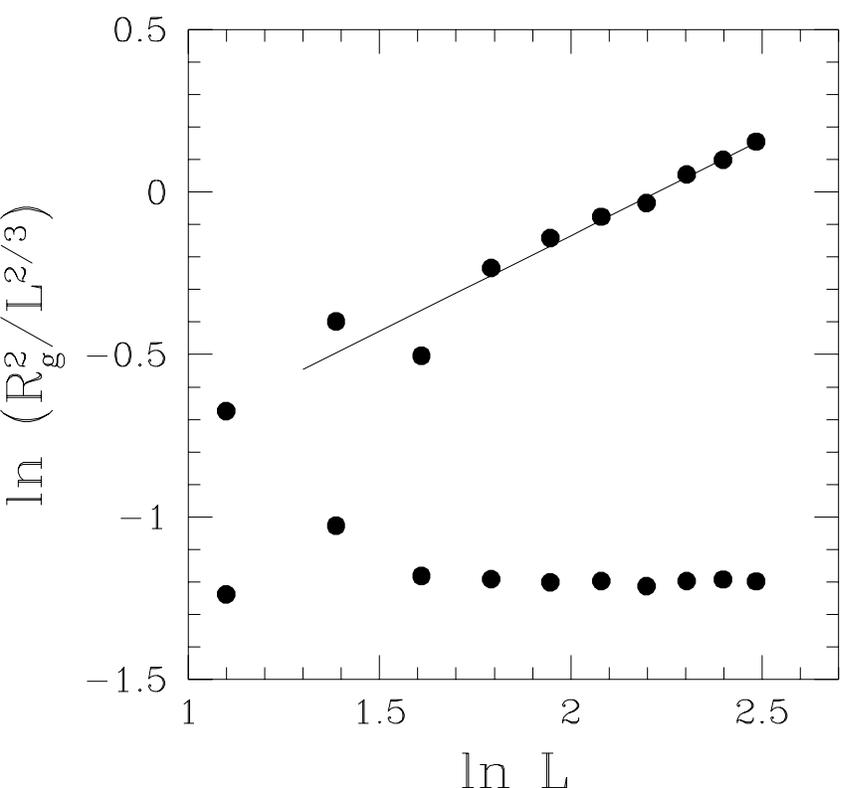
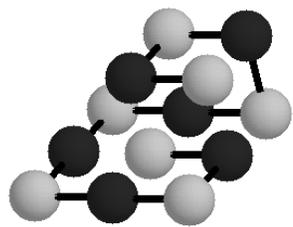
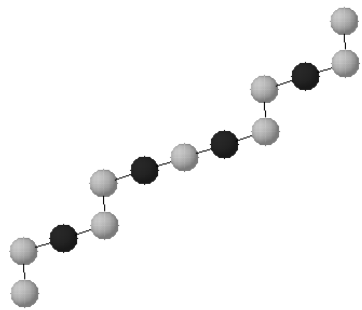
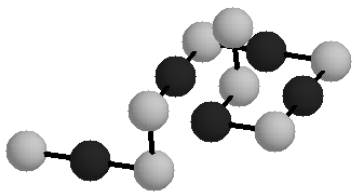
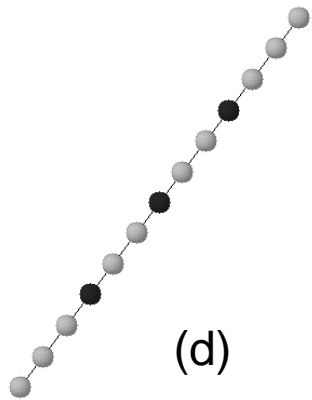
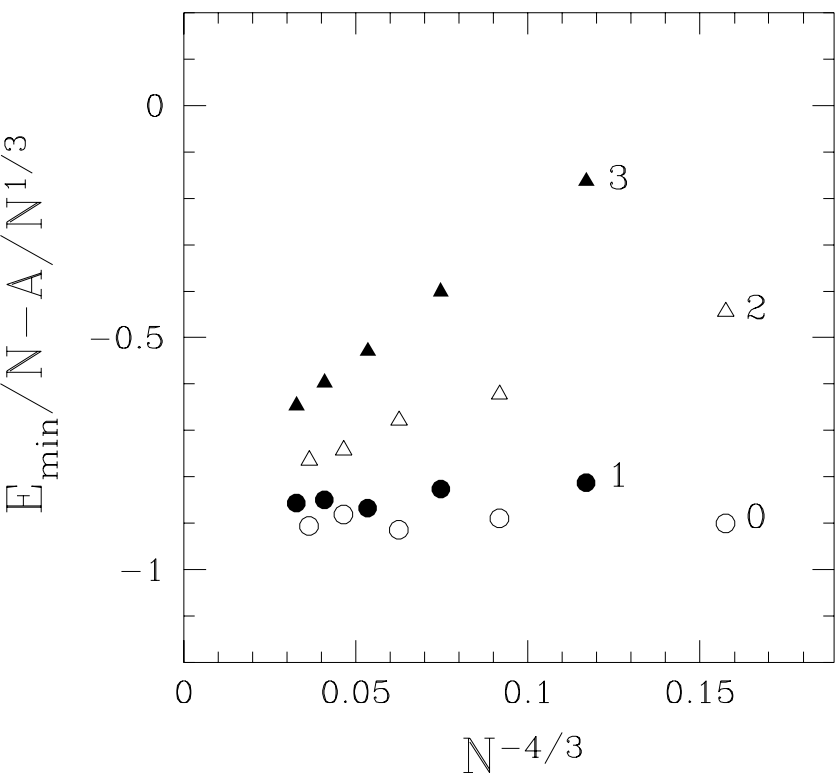
\end{document}